\documentclass[%
 reprint,
 amsmath,amssymb,
 aps,
]{revtex4-2}

\usepackage{graphicx}
\usepackage{hyperref}
\usepackage{physics}
\usepackage[caption=false]{subfig}
\usepackage{booktabs}
\usepackage{dcolumn}
\usepackage{esint}
\usepackage{xcolor}

\immediate\write18{texcount -inc -incbib
-sum borra.tex > /tmp/wordcount.tex}

\immediate\write18{texcount -char -freq
 borra.tex > /tmp/charcount.tex}

\begin{document}

\preprint{APS/123-QED}

\title{Numerical Framework for Modeling Quantum Electromagnetic Systems \\ Involving Finite-Sized Lossy Dielectric Objects in Free Space}
\author{Dong-Yeop Na}
\email{dyna22@postech.ac.kr}
\affiliation{
Department of Electrical Engineering, Pohang University of Science and Technology, Pohang 37673, South Korea\\
}%

\author{Thomas E Roth}%
\affiliation{%
Elmore Family School of Electrical and Computer Engineering, Purdue University, West Lafayette, IN 47907, USA\\
}%

\author{Jie Zhu}%
\affiliation{%
Elmore Family School of Electrical and Computer Engineering, Purdue University, West Lafayette, IN 47907, USA\\
}%

\author{Christopher J Ryu}%
\affiliation{%
Department of Electrical and Computer Engineering, University of Illinois Urbana-Champaign Urbana, IL 61801 USA\\
}%

\author{Weng C. Chew}
\email{wcchew@purdue.edu; corresponding author}
\affiliation{%
Elmore Family School of Electrical and Computer Engineering, Purdue University, West Lafayette, IN 47907, USA\\
}%

\date{\today}

\begin{abstract}
The modified Langevin noise formalism \cite{Drezet2017Quantizing,Stefano2001Mode} has been proposed for the correct charaterization of quantum electromagnetic fields in the presence of finite-sized lossy dielectric objects in free space.
The main modification to the original one \cite{Gruner1996Green,Dung1998three} (also known as the Green's function approach available only for bulk inhomogeneous lossy dielectric medium) was to add fluctuating sources in reaction to the radiation loss.
Consequently, a resulting electric field operator is now determined by (i) boundary-assisted and (ii) medium-assisted fields on an equal footing, which are fluctuating sources due to radiation and medium losses, respectively.
However, due to the lengthy mathematical manipulation and complicated concepts, the validity of the modified Langevin noise formalism has not been clearly checked yet.

In this work, we propose and develop a novel numerical framework for the modified Langevin noise formalism by exploiting computational electromagnetic methods (CEM).
Specifically, we utilize the finite-element method to numerically solve plane-wave-scattering and point-source-radiation problems whose solutions are boundary-assisted and medium-assisted fields, respectively.
Based on the developed numerical framework, we calculate the Purcell factor of a two-level atom inside or outside a lossy dielectric slab.
It is numerically proved, for the first time, that one can retrieve the conventional expression of the spontaneous emission rate, viz., the imaginary part of the Green's function.

The proposed numerical framework is particularly useful for estimating the dynamics of multi-level atoms near practical plasmonic structures or metasurfaces.
\end{abstract}

\maketitle

\section{Introduction}
Handling quantum electromagnetic systems involving a lossy dielectric object in free space (or with open boundary conditions) is challenging. 
This is because of the non-Hermiticity caused by radiation and medium losses.
As a result, the most fundamental properties in quantum physics, for example, equal-time commutator relations for conjugate variables \cite{Ryu2022Fourier}, may not be preserved.
Furthermore, one cannot find eigenmodes with real eigenfrequencies and nice orthonormal properties since a generalized Hermitian eigenvalue problem cannot be derived from such non-Hermitian EM systems.
Consequently, it is not straightforward to apply the classical phenomenological electromagnetic (EM) theory on the standard second quantization procedure unlike lossless cases, e.g., inhomogeneous or anisotropic media \cite{Knoll1987Action,Glauber1991quantum,Chew2016quantum}.

To resolving this critical issue, the microscopic model based on the rigorous Hamiltonian description has been first shown by Huttner and Barnett \cite{Huttner1992Quantization} in an attempt to model a lossy bulk dielectric medium, and a number of subsequent variants \cite{PhysRevA.70.013816,Philbin2010canonical} have been also proposed for the sake of extending the prototype work into more generic cases including medium inhomogeneity and magnetic polarization effects. 
The core idea behind the microscopic model is to introduce the infinite number of harmonic oscillators, called bath oscillators, at every single point in the medium region and accounts for couplings between vacuum EM fields and bath oscillators.
Such couplings could explain mechanisms of the EM energy loss as well as predict the existence of Langevin noise current sources, i.e., fluctuations to the medium loss, while the whole system still remaining to be Hermitian.
Thus, the microscopic model is quantizable in principle, for example, most previous works diagonalized the total Hamiltonian, composed of dynamical variables associated with vacuum EM fields and bath oscillators, utilizing the Fano diagonalization method.
Notably, two recent works \cite{Jauslin2019Canonical,Na2021Diagonalization} have shown exact diagonalization methods in the momentum and position spaces, respectively, which are more suitable for numerical methods; hence, large-scale numerical simulations could be performed. 
However, the numerical diagonalization requires tremendous computational costs, especially, for the real bath which has the infinite degrees of freedom (DoFs) over both space and frequency.  
To deal with the infinite DoFs of bath oscillators, one may take the coarse-graining strategy.\footnote{The coarse-graining technique is widely used in computational plasma science, more specifically the ``Particle-in-Cell'' algorithm \cite{Dawson1983Particle,PINTO2014Charge,NA2017Axisymmetric}, e.g., coarse-graining few millions of actual charged particles into a single superparticle (or computational particle) over the phase space.} 
But the resulting computation expenses are still costly to avoid or delay the Poincar\`e recurrence; otherwise, incorrect energy feedback from bath systems may alter the actual physics of the EM energy loss and fluctuation.

As a great alternative to the microscopic model, a new formalism, called the (previous) Langevin noise model, was proposed by Welsch and coworkers \cite{Gruner1996Green,Dung1998three} based on the fluctuation-dissipation theorem (FDT), being computationally much more efficient since it only keeps track of the EM dynamics by tracing out the dynamics of infinite bath oscillators. 
According to the original Langevin noise formalism, a monochromatic electric field operator is entirely determined by Langevin noise current source operators, taking the form of
\begin{flalign}
\hat{\mathbf{E}}(\mathbf{r},\omega)
=
i\omega\mu_0
\iiint_{V_{m}}d\mathbf{r}'
\overline{\mathbf{G}}(\mathbf{r},\mathbf{r}',\omega)
\cdot
\hat{\mathbf{J}}_{N}(\mathbf{r}',\omega)
\label{eqn:prev_LN}
\end{flalign}
where $V_{m}$ is a volume of lossy dielectric objects described by $\epsilon(\mathbf{r},\omega)$, $\overline{\mathbf{G}}(\mathbf{r},\mathbf{r}',\omega)$ is a dyadic Green's function in the presence of the lossy dielectric objects, and $\hat{\mathbf{J}}_{N}(\mathbf{r}',\omega)$ is a Langevin noise current operator given by
\begin{flalign}
\hat{\mathbf{J}}_{N}(\mathbf{r}',\omega)
=
\frac{\omega}{\mu_0 c^2}\sqrt{\frac{\hbar}{\pi\epsilon_0}\chi_{I}(\mathbf{r}',\omega)}
\hat{\mathbf{f}}(\mathbf{r}',\omega)
\label{eqn:Langevin_Noise_Current_Operator_PLN}
\end{flalign}
for $\mathbf{r}'\in V_{m}$ where $\chi_{I}(\mathbf{r},\omega)$ denotes the imaginary part of the electric susceptibility of the lossy dielectric medium.
In other words, electric fields are supported by fluctuations which are in reaction to medium losses; thus, we can call these ``medium-assisted'' fields.
Vectorial bosonic ladder operators $\hat{\mathbf{f}}(\mathbf{r}',\omega)$ and $\hat{\mathbf{f}}^{\dag}(\mathbf{r}',\omega)$ in \eqref{eqn:Langevin_Noise_Current_Operator_PLN} diagonalize the Hamiltonian operator by
\begin{flalign}
\hat{H}=\int_{0}^{\infty}d\omega\iiint_{V_{m}}d\mathbf{r}'
\hbar\omega\hat{\mathbf{f}}^{\dag}(\mathbf{r}',\omega)\cdot\hat{\mathbf{f}}(\mathbf{r}',\omega).
\nonumber
\end{flalign}
Note that in the above we excluded the zero-point energy for simplicity, and the same will apply to what follows throughout the manuscript.

However, it was argued by two works by Drezet \cite{Drezet2017Quantizing} and Stefano \cite{Stefano2001Mode} that the previous LN model may be an incomplete theory since it omitted the influence of fluctuations reacting to radiation losses, which can be thought of as thermal radiations coming from the infinite boundary $S_{\infty}$.
This missing contribution in the original Langevin noise formalism obviously gets more important when it comes to finite-sized lossy dielectric objects, which represent typical layouts of optical components. 
The modified Langevin noise formalism added the missing term (we shall call this ``boundary-assisted'' fields) into the previous LN model, shall be discussed in detail later. 
As such, the modified Langevin noise formalism can fully agree with the FDT's argument that the EM dynamics is now determined by two different fluctuations in reaction to radiation and medium losses.
Especially, the modified Langevin noise formalism would be useful in practical quantum optics problems, .e.g, studying and engineering quantum plasmonic devices or metasurface-based quantum information science technologies.
To do this, one should be able to evaluate both boundary-assisted and medium-assisted fields in the presence of arbitrary lossy dielectric objects including geometric complexity and medium inhomogenity.
However, their closed-form solutions are limited to very simple cases and unvailable for most cases.

Motivated by this, we have dedicated over the past years into building numerical frameworks for modeling quantum optics and circuit quantum electrodynamics (QED) phenomena based on classical computational electromagnetics (CEM) methods by reinterpreting and refining the existing math-physics models using various CEM methods \cite{Na2020quantum,Na2020QuantumFD,roth_2020,Roth2021Macroscopic,Xia2021Casimir,Chew2021Quantum,Na2021Diagonalization}.
We believe that our efforts of transplanting CEM methods into quantum physics will become a stepping stone to further advance the research paradigm in the existing quantum technology, which mostly relies on theory and experiments, and accelerate the realization of quantum science and technology.

In this article, we present a numerical framework for quantitative analyses on quantum EM systems including lossy dielectric objects with the open boundary by incorporating computational electromagnetic methods (CEM) into the modified Langevin noise formalism.
To our knowledge, however, no previous works exist yet that incorporate numerical methods into the modified Langevin noise formalism and performing fully-quantum-theoretic numerical simulations.
We shall discuss about the modified Langevin noise formalism in detail how the electric field operator can be determined by both boundary-assisted (BA) and medium-assisted (MA) fields on an equal footing, which result from fluctuations in reaction to radiation and medium losses.
Numerical solutions to BA and MA fields are found based on the finite-element method that solves standard plane-wave-scattering problems and point-source-radiation problems, respectively.
Especially, we connect the modified Langevin noise formalism to the spectral function (SFA) \cite{Chew2019Green}, deriving the thermal equilibrium condition from the use of the correct dyadic-dyadic Green theorem \cite{Drezet2017Equivalence} to show that BA/MA fields can make open and lossy EM systems quasi-Hermitian or in the thermal equilibrium.
Finally, we shall consider a numerical example of Purcell factors of a two-level system located inside or outside a lossy dielectric slab.
We compare the calculation results obtained by various methods, such as, the numerical diagonalization method \cite{Dorier2019Canonical,Na2021Diagonalization} for the microscopic model \cite{Philbin2010canonical,Sha2018Dissipative}, SFA, previous LN model, and modified Langevin noise formalism.

The contributions of the present work are twofold:
\begin{itemize}
\item We build a new numerical framework for analyzing quantum optics problems involving the radiation and medium losses by incorporating the use of computational electromagnetic methods into the modified Langevin noise formalism. Especially, we provide specific numerical recipes in solving plane-wave-scattering and point-source-radiation problems. The formal and later are of boundary-assisted and medium-assisted fields, respectively. 

\item With the use of developed numerical framework, we numerically prove that the use of modified Langevin noise formalism can retrieve the conventional expression of the spontaneous emission rate of a two-level atom inside or outside a lossy dielectric object(s), viz., the imaginary part of the Green's function.
\end{itemize}
It implies that when analyzing interactions between atoms and EM fields around plasmonic nano-particles or structures, one has to consider effects of BA fields as well as MA fields.
But most of previous works only considered MA effects.

The paper is organized as follows. 
Sec. II presents the essence and main features of the modified Langevin noise formalism are presented. 
Specifically, it is shown that BA/MA fields, which are the main ingredients of the theory, can be found from plane-wave-scattering and point-source-radiation problems.
Sec. III presents the detailed numerical recipe to solve the plane-wave-scattering and point-source-radiation problems in modeling BA/MA fields.
In particular we utilize the finite element method in frequency domain with the use of perfectly matched layers to model the radiation loss properly.
In Sec. IV the modified Langevin noise formalism is connected to the spectral function approach by considering the field correlation.
With the proper use of the dyadic-dyadic Green's function, we hypothsize the thermal equilibrium identity which is to be examined in Sec. V.
In Sec. V one-dimensional simulation results of Purcell factors of a two-level atom inside or outside a lossy dielectric slab are discussed.
Calculations based on the Fermi-Golden rule were performed by using four different methods: (i) the spectral function approach, i.e., the imaginary part of the Green's function, (ii) the second quantization for the microscopic model via numerical mode decomposition, (iii) the modified Langevin noise formalism, and (iv) the original Langevin noise formalism.
A summary and conclusions are given in Sec. VI.
The operator-form dyadic-dyadic Green theorem is discussed in detail in Appendix A.

\section{Modified Langevin noise formalism: Boundary- and Medium-Assisted fields}\label{sec_II}
\begin{figure*}
\centering
\includegraphics[width=0.8\linewidth]{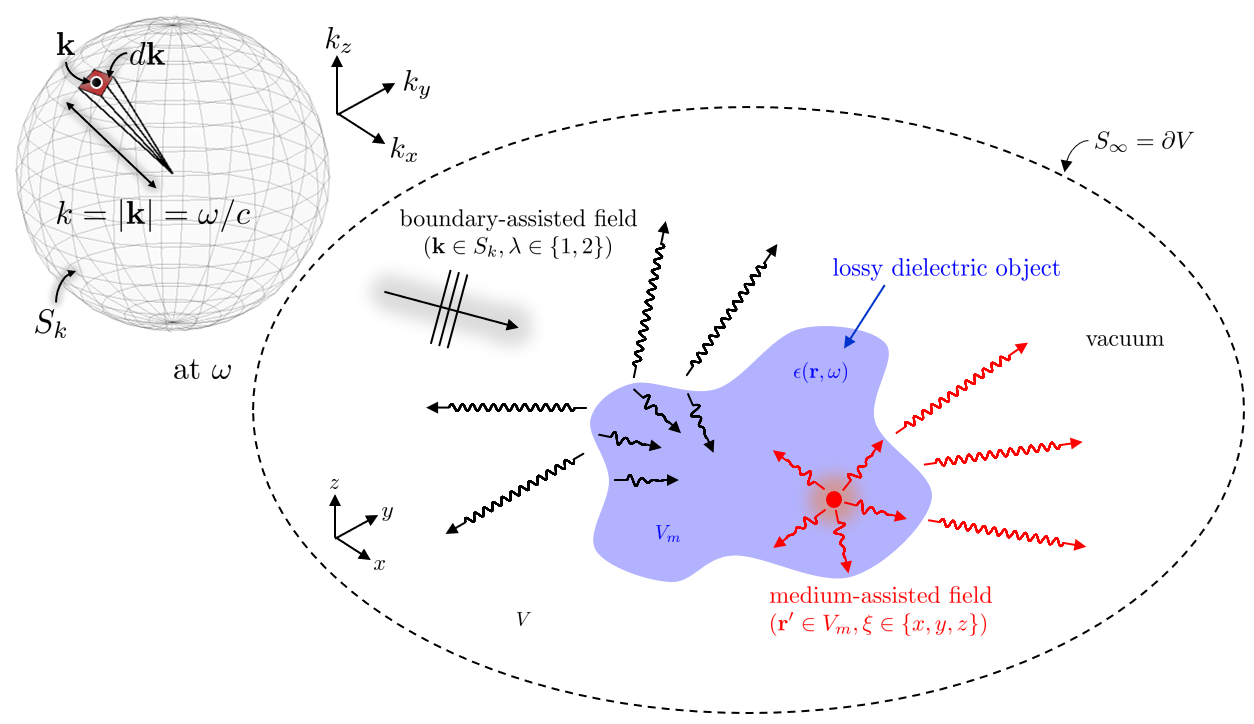}
\caption{
Illustration of monochromatic boundary-assisted and medium-assisted (BA/MA) fields. 
Each degenerate BA field is a total field composed of an incident plane wave with $(\mathbf{k} \in S_{k},\lambda \in \left\{1,2\right\})$ and resulting scattered fields by a lossy dielectric object.
On the other hand, each degenerate MA field is a radiating field by a point current source $(\mathbf{r}'\in V_{m},\xi\in \left\{x,y,z\right\})$ embedded in the lossy dielectric object.
}
\label{fig:BA/MA_schematic}
\end{figure*} 
Consider a lossy dielectric (non-magnetic) object in the vacuum background, as illustrated in Fig. \ref{fig:BA/MA_schematic}. 
The effective permittivity of the lossy dielectrib object is given by
\begin{flalign}
&\epsilon(\mathbf{r},\omega)
=
\epsilon_0 \epsilon_r(\mathbf{r},\omega)
=
\epsilon_0\Bigl(1+\chi(\mathbf{r},\omega)\Bigr)
\nonumber \\
&\quad
=
\left\{
\begin{matrix}
\epsilon_0\Bigl(1+\chi_{R}(\mathbf{r},\omega)+i\chi_{I}(\mathbf{r},\omega)\Bigr), & \text{for~} \mathbf{r}\in V_{m}
\\
\epsilon_0, & \text{elsewhere}
\end{matrix}
\right.
,
\label{eqn:effective_permittivity}
\end{flalign}
where $V_{m}$ is the volume of the lossy dielectric object.
The lossy dielectric object is assumed to be causal while satisfying the Kramers-Kronig relation.

According to the modified Langevin noise formalism \cite{Drezet2017Quantizing,Stefano2001Mode}, the complete solution to a monochromatic electric field operator should include boundary-assisted (BA) and medium-assisted (MA) fields, its positive-frequency part taking the form of
\begin{flalign}
\hat{\mathbf{E}}(\mathbf{r},\omega)
=
\hat{\mathbf{E}}_{(\text{B})}(\mathbf{r},\omega)
+
\hat{\mathbf{E}}_{(\text{M})}(\mathbf{r},\omega)
\label{eqn:BA/MA_fields}
\end{flalign}
where
\begin{flalign}
&\hat{\mathbf{E}}_{(\text{B})}(\mathbf{r},\omega)
=
i
\frac{1}{\left(\sqrt{2\pi}\right)^{3}}
\oiint_{S_{k}} d\mathbf{k}
\sum_{\lambda\in\left\{1,2\right\}}
\nonumber \\
&\quad\quad\quad\quad\quad
\boldsymbol{\Phi}_{(\text{tot})}(\mathbf{r},\mathbf{k},\lambda,\omega)
\sqrt{\frac{\hbar\omega}{2}}
\hat{a}(\mathbf{k},\lambda,\omega),
\label{eqn:BA_field}
\\
&\hat{\mathbf{E}}_{(\text{M})}(\mathbf{r},\omega)
=
i\frac{\omega^2}{c^2}
\iiint_{V_{m}}d\mathbf{r}'
\sum_{\xi\in\left\{x,y,z\right\}}
\nonumber \\
&\quad\quad\quad\quad\quad
\left(
\overline{\mathbf{G}}(\mathbf{r},\mathbf{r}',\omega)
\cdot
\hat{\xi}
\right)
\sqrt{\frac{\hbar\chi_{I}(\mathbf{r}',\omega)}{\pi\epsilon_0}}
\hat{f}(\mathbf{r}',\xi,\omega).
\label{eqn:MA_field}
\end{flalign}
In the above, $\mathbf{k}$ is a wavevector, $k=\left|\mathbf{k}\right|=\omega/c$ is the wavenumber, $S_{k}$ is the surface of the radiation sphere in $\mathbf{k}$-space, and $\lambda$ denotes the polarization degeneracy index for an incident plane wave coming from the infinity $S_{\infty}$.
Note that $\overline{\mathbf{G}}(\mathbf{r},\mathbf{r}',\omega)$ is the dyadic Green's function in the presence of the lossy dielectric object from which
\begin{flalign}
\Bigl(\nabla\times\nabla\times
-\frac{\omega^2}{c^2}\epsilon_r(\mathbf{r},\omega)
\Bigr)\overline{\mathbf{G}}(\mathbf{r},\mathbf{r}',\omega)
=
\overline{\mathbf{I}}\delta(\mathbf{r}-\mathbf{r}')
\label{eqn:DGF}
\end{flalign}
for $\mathbf{r}'\in V_{m}$.
It should be pointed out that \eqref{eqn:MA_field} is the same as \eqref{eqn:prev_LN} while explicitly writing the Langevin noise current source operator in \eqref{eqn:Langevin_Noise_Current_Operator_PLN} in terms of $\hat{f}(\mathbf{r}',\xi,\omega)=\hat{\xi}\cdot\hat{\mathbf{f}}(\mathbf{r}',\omega)$ for $\xi \in \left\{x,y,z\right\}$.

One can observe two important properties from the modified Langevin noise formalism with BA/MA fields in \eqref{eqn:BA/MA_fields}: (i) a monochromatic electric field operator is expanded by the infinite number of degenerate BA and MA fields originating from two fluctuation sources due to radiation and medium losses, respectively, and (ii) the degeneracy indices of BA/MA fields are descended from original degrees of freedom (DoFs) for (vacuum) photonic systems and reservoir oscillator fields, i.e., BA fields take the degeneracy in terms of ($\mathbf{k}\in S_{k},\lambda$) same as that of plane waves in the vacuum, and MA fields form the degeneracy with respect to ($\mathbf{r}\in V_{m},\xi$) which are of bath oscillators.

\subsection{Boundary-assisted fields}
One can notice from \eqref{eqn:BA_field} that the monochromatic BA field is expanded by many different BA fields in terms of $\mathbf{k}$ and $\lambda$.
Hence, $(\mathbf{k},\lambda)$ can be thought of as sort of the degeneracy index of BA fields.
Then, each degenerate BA field having $(\mathbf{k},\lambda)$ corresponds to a total field $\boldsymbol{\Phi}_{\text{tot}}(\mathbf{r},\mathbf{k},\lambda,\omega)$ consisting of (i) an incident plane wave with ($\mathbf{k}\in S_{k}$, $\lambda\in \left\{1,2\right\}$) and (ii) resulting scattered fields by the lossy dielectric object; hence, 
\begin{flalign}
\boldsymbol{\Phi}_{(\text{tot})}(\mathbf{r},\mathbf{k},\lambda,\omega)
&=
\boldsymbol{\Phi}_{(\text{inc})}(\mathbf{r},\mathbf{k},\lambda,\omega)
\nonumber \\
&+
\boldsymbol{\Phi}_{(\text{sca})}(\mathbf{r},\mathbf{k},\lambda,\omega),
\label{eqn:BA_total_field}
\end{flalign}
satisfying 
\begin{flalign}
\Bigl(\nabla\times\nabla\times
-\frac{\omega^2}{c^2}\epsilon_r(\mathbf{r},\omega)
\Bigr)
\boldsymbol{\Phi}_{(\text{tot})}(\mathbf{r},\mathbf{k},\lambda,\omega)
=
0.
\label{eqn:homogen_sol_eqn}
\end{flalign}
When substituting \eqref{eqn:BA_total_field} into \eqref{eqn:homogen_sol_eqn} with the use of \eqref{eqn:effective_permittivity}, one can arrive at the following plane-wave-scattering problem:
\begin{flalign}
&
\Bigl(\nabla\times\nabla\times
-\frac{\omega^2}{c^2}\epsilon_r(\mathbf{r},\omega)
\Bigr)
\boldsymbol{\Phi}_{(\text{sca})}(\mathbf{r},\mathbf{k},\lambda,\omega)
\nonumber \\
&\quad\quad\quad\quad\quad\quad\quad
=
\frac{\omega^2}{c^2}\chi(\mathbf{r},\omega)
\boldsymbol{\Phi}_{(\text{inc})}(\mathbf{r},\mathbf{k},\lambda,\omega)
\label{eqn:plane_wave_scattering_problem}
\end{flalign}
since 
\begin{flalign}
\Bigl(\nabla\times\nabla\times
-\frac{\omega^2}{c^2}
\Bigr)
\boldsymbol{\Phi}_{(\text{inc})}(\mathbf{r},\mathbf{k},\lambda,\omega)
=0.
\end{flalign}
Since the incident plane wave is known, for example, $\boldsymbol{\Phi}_{(\text{inc})}(\mathbf{r},\mathbf{k},\lambda,\omega)=\hat{e}_{\lambda}e^{i\mathbf{k}\cdot\mathbf{r}}$ where $\hat{e}_{\lambda}$ is a polarization unit vector, one can solve \eqref{eqn:plane_wave_scattering_problem} for the scattered fields $\boldsymbol{\Phi}_{(\text{sca})}(\mathbf{r},\mathbf{k},\lambda,\omega)$. 

\subsection{Hamiltonian operator diagonalized by $\hat{a}$ and $\hat{f}$}
The monochromatic Hamiltonian operator in the modified Langevin noise formalism is expressible in terms of two different diagonalizing ladder operators $\hat{a}$ and $\hat{f}$:
\begin{flalign}
\hat{H}(\omega)
&=
\oiint_{S_{k}} d\mathbf{k} 
\sum_{\lambda\in\left\{1,2\right\}}
\hbar\omega
\hat{a}^{\dag}(\mathbf{k},\lambda,\omega)
\hat{a}(\mathbf{k},\lambda,\omega)
\nonumber \\
&+
\iiint_{V_{m}}d\mathbf{r}'
\sum_{\xi\in\left\{x,y,z\right\}}
\hbar \omega \hat{f}^{\dag}(\mathbf{r}',\xi,\omega)\hat{f}(\mathbf{r}',\xi,\omega).
\end{flalign}
It should be mentioned that the above Hamiltonian does not represent the EM energy only.

 but describes total energy of the whole system.
Here, the whole system refers to EM systems plus two thermal baths.
This is because we introduced the medium and radiation losses, there should be two different thermal baths for the losses, respectively. 
And the EM systems are in thermal equilibrium with these thermal baths. 
Hence, the physical meaning of the above Hamiltonian operator is total energy contained in the EM system as well as the two thermal baths.
The Hamiltonian is then constant of motion; hence, it is energy conserving.
Furthermore, the Hamiltonian operator is diagonalized by ladder operators $\hat{a}$ and $\hat{f}$ associated with the medium and radiation fluctuations.

Eigenstates for the above Hamiltonian operator are two different kinds of Fock states associated with BA/MA fields, i.e.,
\begin{flalign}
\hat{n}(\mathbf{k},\lambda,\omega)\ket{n}_{\mathbf{k},\lambda,\omega}
&=
n\ket{n}_{\mathbf{k},\lambda,\omega},\\
\hat{m}(\mathbf{r}',\xi,\omega)\ket{m}_{\mathbf{r}',\xi,\omega}
&=
m\ket{m}_{\mathbf{r}',\xi,\omega}.
\end{flalign}
In the above, $\hat{n}(\mathbf{k},\lambda,\omega)=\hat{a}^{\dag}(\mathbf{k},\lambda,\omega)\hat{a}(\mathbf{k},\lambda,\omega)$ is number operator , $\ket{n}_{\mathbf{k},\lambda,\omega}$ is Fock state, and $n$ is the number of quanta for a BA field having $(\mathbf{k},\lambda)$.
On the other hand, $\hat{m}(\mathbf{r}',\xi,\omega)=\hat{f}^{\dag}(\mathbf{r}',\xi,\omega)\hat{f}(\mathbf{r}',\xi,\omega)$ is number operator, $\ket{m}_{\mathbf{k},\lambda,\omega}$ is Fock state, and $m$ is the number of quanta for a MA field specified by $(\mathbf{r}',\xi)$.

The Fock states satisfy the following orthonormal properties:
\begin{flalign}
\bra{n'}_{ {\mathbf{k}'},{\lambda'},\omega' }
\ket{n}_{ \mathbf{k},\lambda,\omega }
&=
\delta_{n',n}
\delta({\mathbf{k}'}-{\mathbf{k}})
\delta_{{\lambda'},{\lambda}}
\delta(\omega'-\omega),
\label{eqn:Ortho_Fock_1}
\\
\bra{m'}_{\mathbf{r}',\xi',\omega'}
\ket{m}_{\mathbf{r},\xi,\omega}
&=
\delta_{m',m}
\delta(\mathbf{r}'-\mathbf{r})
\delta_{\xi',\xi}
\delta(\omega'-\omega),
\label{eqn:Ortho_Fock_2}
\\
\bra{n'}_{ {\mathbf{k}},{\lambda},\omega}
\ket{m}_{\mathbf{r},\xi,\omega}
&=0.
\label{eqn:Ortho_Fock_3}
\end{flalign}
The action of the bosonic ladder operators on Fock states can be evaluated by
\begin{flalign}
\hat{a}(\mathbf{k},\lambda,\omega)\ket{n}_{\mathbf{k},\lambda,\omega}
&=
\sqrt{n}\ket{n-1}_{\mathbf{k},\lambda,\omega},
\\
\hat{a}^{\dag}(\mathbf{k},\lambda,\omega)\ket{n}_{\mathbf{k},\lambda,\omega}
&=
\sqrt{n+1}\ket{n+1}_{\mathbf{k},\lambda,\omega},
\\
\hat{f}(\mathbf{r}',\xi,\omega)\ket{m}_{\mathbf{r}',\xi,\omega}
&=
\sqrt{m}\ket{m-1}_{\mathbf{r}',\xi,\omega},
\\
\hat{f}^{\dag}(\mathbf{r}',\xi,\omega)\ket{m}_{\mathbf{r}',\xi,\omega}
&=
\sqrt{m+1}\ket{m+1}_{\mathbf{r}',\xi,\omega}.
\end{flalign}

The diagonalizing ladder operators are supposed to satisfy the following standard bosonic commutator relations:
\begin{flalign}
&\left[
\hat{a}(\mathbf{k},\lambda,\omega),
\hat{a}^{\dag}(\mathbf{k}',\lambda',\omega')
\right]
=\hat{I}
\delta(\mathbf{k}-\mathbf{k}')
\delta_{\lambda,\lambda'}
\delta(\omega-\omega'),
\nonumber \\
&\left[
\hat{a}(\mathbf{k},\lambda,\omega),
\hat{a}(\mathbf{k}',\lambda',\omega')
\right]
=
0
\nonumber \\
&\quad\quad\quad\quad\quad\quad\quad\quad\quad
=
\left[
\hat{a}^{\dag}(\mathbf{k},\lambda,\omega),
\hat{a}^{\dag}(\mathbf{k}',\lambda',\omega')
\right],
\label{eqn:BC_a}
\\
&\left[
\hat{f}(\mathbf{r},\xi,\omega),
\hat{f}^{\dag}(\mathbf{r}',\xi',\omega'),
\right]
=\hat{I}
\delta(\mathbf{r}-\mathbf{r}')
\delta_{\xi,\xi'}
\delta(\omega-\omega'),
\nonumber \\
&\left[
\hat{f}(\mathbf{r},\xi,\omega),
\hat{f}(\mathbf{r}',\xi',\omega')
\right]
=
0
\nonumber \\
&\quad\quad\quad\quad\quad\quad\quad\quad\quad
=
\left[
\hat{f}^{\dag}(\mathbf{r},\xi,\omega),
\hat{f}^{\dag}(\mathbf{r}',\xi',\omega')
\right].
\label{eqn:BC_f}
\end{flalign}

The total Hamiltonian and electric field operators are then obtained by integrating the monochromatic terms over the frequency domain:
\begin{flalign}
\hat{H}
&=
\int_{0}^{\infty} d\omega \hat{H}(\omega),
\\
\hat{\mathbf{E}}(\mathbf{r},t)
&=
\int_{0}^{\infty}d\omega
\hat{\mathbf{E}}(\mathbf{r},\omega)e^{-i\omega t}
+
\text{h.c.}
\label{eqn:full_E_operator}
\end{flalign}

\section{Numerical solutions of BA/MA fields Using Finite Element Method}\label{sec_V}
Here, we provide numerical recipes to find approximate solutions of BA/MA fields based on the finite element method (FEM) \cite{Jin2002-fl}.
BA/MA fields can be found by solving (i) plane-wave-scattering problems and (ii) point-source-radiation problems, respectively.

\subsection{Plane-wave-scattering problems for BA fields}
\begin{figure}
\centering
\includegraphics[width=\linewidth]{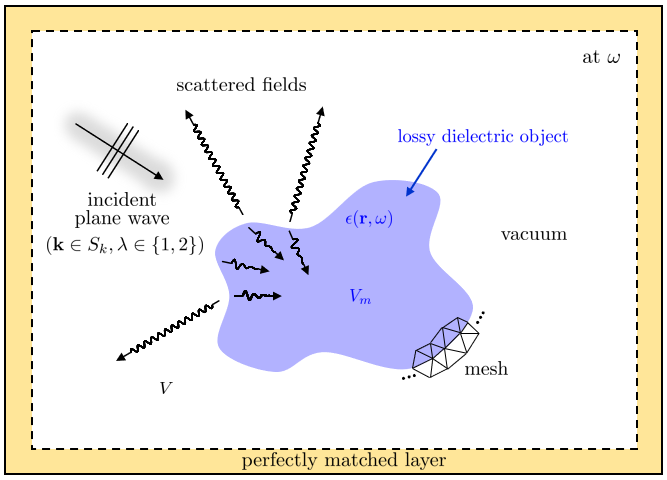}
\caption{Finding a numerical solution of each degenerate BA field per one FEM simulation modeling a plane-wave-scattering problem.}
\label{fig:BA_FEM}
\end{figure}

Consider a lossy dielectric object in the vacuum background, and assume that a plane wave with ($\mathbf{k}\in S_{k}$, $\lambda\in\left\{1,2\right\}$) at $\omega$ is incident on the scatterer, as illustrated in Fig. \ref{fig:BA_FEM}.

One needs to first prepare unstructured meshes that reconstruct the original problem geometry.
Unknown scattered fields are then expanded by Whitney 1-forms (or edge elements) related to edges of the mesh, such as
\begin{flalign}
\boldsymbol{\Phi}_{(\text{sca})}(\mathbf{r},\mathbf{k},\lambda,\omega)
\approx
\sum_{i=1}^{N_{1}}
\left[\boldsymbol{\varphi}^{(\text{sca})}_{\mathbf{k},\lambda,\omega}\right]_{i}
\mathbf{W}_{i}^{(1)}(\mathbf{r}).
\label{eqn:scattered_field_W1}
\end{flalign}
where $N_1$ is the number of edges, $\boldsymbol{\varphi}^{(\text{sca})}_{\mathbf{k},\lambda,\omega}$ is a one-dimensional vector array listing degrees of freedom for the scattered fields, and $\mathbf{W}_{i}^{(1)}(\mathbf{r})$ denotes the Whitney 1-form for $i$-th edge.
Substituting \eqref{eqn:scattered_field_W1} into \eqref{eqn:plane_wave_scattering_problem} and performing the Galerkin testing, one can find the following linear system which is the discrete counterpart of \eqref{eqn:plane_wave_scattering_problem} expressed by
\begin{flalign}
\Bigl(
\overline{\mathbf{S}}
-
\frac{\omega^2}{c^2}\overline{\mathbf{M}}
\Bigr)
\cdot \boldsymbol{\varphi}^{(\text{sca})}_{\mathbf{k},\lambda,\omega}
=
\mathbf{f}^{(\text{inc})}_{\mathbf{k},\lambda,\omega}
\label{eqn:incident-and-scattered_field_FEM_solver}
\end{flalign}
where $\overline{\mathbf{S}}$ and $\overline{\mathbf{M}}$ denote stiffness and mass matrices that encode $\nabla\times\mu^{-1}_{0}\nabla\times$ and $\epsilon(\mathbf{r},\omega)$, respectively. 
And $\mathbf{f}^{(\text{inc})}_{\mathbf{k},\lambda,\omega}$ is a force vector whose $i$\textsuperscript{th} element can be evaluated by
\begin{flalign}
\left[\mathbf{f}^{(\text{inc})}_{\mathbf{k},\lambda,\omega}\right]_{i}
=
\left<\mathbf{W}^{(1)}_{i}(\mathbf{r}),
\frac{\omega^2}{c^2}\chi(\mathbf{r},\omega)
\boldsymbol{\Phi}_{(\text{inc})}(\mathbf{r},\mathbf{k},\lambda,\omega)
\right>
\end{flalign}
where $\left<\mathbf{A},\mathbf{B}\right>$ denotes the projection process, i.e., the spatial integral of the inner product of two vector fields $\mathbf{A}$ and $\mathbf{B}$ over a certain finite support.
Solving \eqref{eqn:incident-and-scattered_field_FEM_solver}, one can find a numerical solution for a degenerate BA field, such as,
\begin{flalign}
\boldsymbol{\Phi}_{(\text{tot})}(\mathbf{r},\mathbf{k},\lambda,\omega)
&\approx
\boldsymbol{\Phi}_{(\text{inc})}(\mathbf{r},\mathbf{k},\lambda,\omega)
\nonumber\\
&+
\sum_{i=1}^{N_{1}}
\left[\boldsymbol{\varphi}^{(\text{sca})}_{\mathbf{k},\lambda,\omega}\right]_{i}
\mathbf{W}_{i}^{(1)}(\mathbf{r}).
\end{flalign}

As mentioned earlier, BA fields originate from the fluctuation to the radiation loss; hence, one should incorporate open boundary conditions in the above FEM simulations so that the radiation loss can be properly taken into account.
Here, we employ perfectly matched layers (PML)\textemdash one kind of absorbing boundary conditions\textemdash based on the complex coordinate stretching method incorporated into PML constitutive tensors \cite{Chew3DPML1994,Teixeira1998General}.
Fig. \ref{fig:BA_FEM} illustrates how to find numerical solutions to BA fields via the FEM simulations.

\subsection{Point-source-radiation problems for MA fields}
Each degenerate MA field can be found by solving a point-source-radiation problem, viz., finding a numerical soluation of the dyadic Green's function whose point source is embedded in the lossy dielectric object.
Taking the numerical dyadic Green's function approach \cite{Gan2017Hybridization} and using the similar FEM implementation with PMLs, one can evaluate numerical dyadic Green's functions by
\begin{flalign}
&\overline{\mathbf{G}}(\mathbf{r},\mathbf{r}',\omega)
\approx
\sum_{i=1}^{N_{1}}
\sum_{j \in \mathbf{j}}
\left[
\overline{\mathbf{L}}^{-1}
\right]_{i,j}
\mathbf{W}_{i}^{(1)}(\mathbf{r})
\otimes
\mathbf{W}_{j}^{(1)}(\mathbf{r}')
\end{flalign}
for point sources located at $\mathbf{r}'\in V_{m}$ where matrix operator $\overline{\mathbf{L}}=\overline{\mathbf{S}}-\omega^2\overline{\mathbf{M}}$, $\otimes$ denotes the tensor product, and $\mathbf{j}$ denotes an integer set whose elements are edge indices of a tetrahedron (3D) or triangle (2D) which include the point source.
\begin{figure}
\centering
\includegraphics[width=\linewidth]{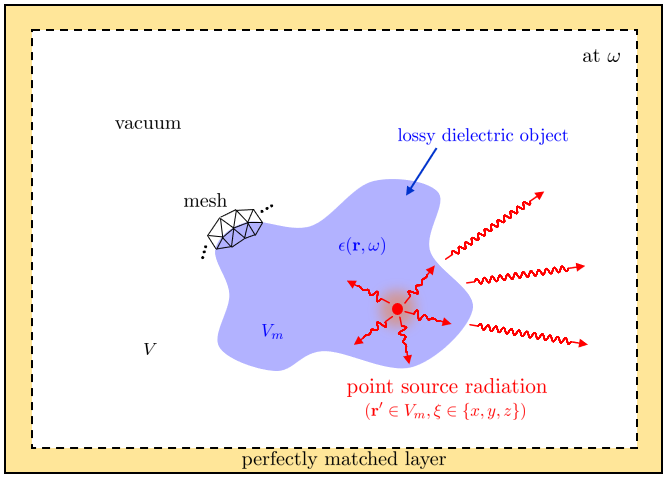}
\caption{Finding a numerical solution of each degenerate MA field per one FEM simulation modeling a point-source-radiation problem.}
\label{fig:MA_FEM}
\end{figure}

\section{Connecting BA/MA Field Correlation to Spectral Function Approach}\label{sec_IV}
\subsection{Spectral function}
The spectral function approach (SFA) is often used to describe the quantum transport in solid-state physics \cite{datta_2005}.
Also, the SFA is widely used to evaluate enhanced spontaneous emission rates of atoms in optics.
Here, we connect the modified Langevin noise formalism to the SFA with the use of the correct dyadic-dyadic Green theorem \cite{Drezet2017Equivalence} to show that BA/MA fields can achieve the thermal equilibrium in open/lossy EM systems.

The spectral function $\overline{\mathbf{A}}_{T}(\mathbf{r},\mathbf{r}',\omega)$ at temperature $T$ is defined to be the sum of retarded and advanced dyadic Green's functions \cite{Chew2019Green}: 
\begin{flalign}
\overline{\mathbf{A}}_{T}(\mathbf{r}_a,\mathbf{r}_b,\omega)
&=
i
\Bigl(
\overline{\mathbf{G}}(\mathbf{r}_a,\mathbf{r}_b,\omega)
-
\overline{\mathbf{G}}^{*}(\mathbf{r}_a,\mathbf{r}_b,\omega)
\Bigr)
\nonumber \\
&=
-2\Im\left\{
\overline{\mathbf{G}}(\mathbf{r}_a,\mathbf{r}_b,\omega)
\right\}.
\label{eqn:spectral_function}
\end{flalign}
The spectral function describes an EM system in thermal equilibrium \cite{Chew2019Green}. 
In a lossy medium, the first term of the spectral function describes a decaying field (retarded), but the second term, a back propagating field, describes a growing field (advanced). 
This can be thought of as a lossy EM system in equilibrium with a thermal bath. 
More specifically, the loss in the EM system is accompanied by Langevin sources induced by the thermal excitation of the environment; hence, the EM system is in thermal equilibrium with the Langevin sources due to the medium's loss. 
On the other hand, in the lossless case with open boundary conditions, e.g., free space, the system is in thermal equilibrium with sources at infinity. 

The spectral function can be also related to the field correlation function $\overline{\mathbf{C}}_{T}(\mathbf{r}_a,\mathbf{r}_b,\omega)$ such a way that 
\begin{flalign}
\overline{\mathbf{A}}_{T}(\mathbf{r}_a,\mathbf{r}_b,\omega)
=
-\frac{\pi}{\omega\mu_0 \Theta(\hbar\omega)}\overline{\mathbf{C}}_{T}(\mathbf{r}_a,\mathbf{r}_b,\omega)
\label{eqn:SFA_FC_rel}
\end{flalign}
where field correlation function is defined to be \cite{Chew2019Green}
\begin{flalign}
\overline{\mathbf{C}}_{T}(\mathbf{r}_a,\mathbf{r}_b,\omega)
&=
\text{Tr}
\left(\hat{\rho}_{\text{th}}
\hat{\mathbf{E}}(\mathbf{r}_a,\omega)
\otimes
\hat{\mathbf{E}}^{\dag}(\mathbf{r}_b,\omega)
\right).
\label{eqn:SF_field_correlation_gen}
\end{flalign}
Note that in the above $\hat{\rho}_{\text{th}}$ denotes a density operator for the thermal state, and the average photon energy density is given by \cite{Chew2019Green}
\begin{flalign}
\Theta(\hbar\omega)=\left(\bar{n}+\frac{1}{2}\right)\hbar\omega
\end{flalign} 
where the average photon number of the thermal field is calculated as \cite{Gerry2004introductory,fox2006quantum}
\begin{flalign}
\bar{n} = \frac{1}{\text{exp}\left(\hbar\omega/k_{B}T\right)-1}.
\end{flalign}
When $T=0$ so $\bar{n}$ approaching zero, the average photon energy density becomes a zero-point energy, i.e., $\Theta(\hbar\omega)=\hbar\omega/2$.
It implies that the quantum harmonic oscillator has nonzero energy even if it is in the ground state at $T=0$.
As a result, the field correlation function at $T=0$ now becomes $\overline{\mathbf{C}}_\text{0}(\mathbf{r}_a,\mathbf{r}_b,\omega)$ as defined in \eqref{eqn:SF_field_correlation}.
And the spectral function at $T=0$ can be related to the quantum field correlator such as
\begin{flalign}
\overline{\mathbf{A}}_{0}(\mathbf{r}_a,\mathbf{r}_b,\omega)
=
-\frac{2\pi}{\hbar\omega^2 \mu_0 }\overline{\mathbf{C}}_{0}(\mathbf{r}_a,\mathbf{r}_b,\omega).
\end{flalign}

\subsection{Field Correlation of BA/MA Fields}
We now consider a quantum field correlator defined by
\begin{flalign}
\overline{\mathbf{C}}_{0}(\mathbf{r}_a,\mathbf{r}_b,\omega)
&=
\mel
{0}
{
\hat{\mathbf{E}}(\mathbf{r}_a,\omega)
\otimes
\hat{\mathbf{E}}^{\dag}(\mathbf{r}_b,\omega)
}
{0}
\label{eqn:SF_field_correlation}
\end{flalign}
where $\ket{0}$ represents the vacuum state.
In what follows, we use $\expval{\hat{O}}$ implicitly representing $\mel{0}{\hat{O}}{0}$ for an operator $\hat{O}$.
(Specifically, when $\mathbf{r}_a=\mathbf{r}_b$, it physically represents the spontaneous emission rate (SER) of a two-level atom in accordance with the Fermi-Golden rule \cite{SCHEEL2008numerical,novotny_hecht_2012}.)

Substituting \eqref{eqn:BA/MA_fields} into \eqref{eqn:SF_field_correlation}, we can rewrite the quantum field correlator in terms of BA/MA fields by
\begin{flalign}
&\left<
\hat{\mathbf{E}}(\mathbf{r}_a,\omega)
\otimes
\hat{\mathbf{E}}^{\dag}(\mathbf{r}_b,\omega)
\right>
\nonumber \\
&\quad\quad\quad\quad\quad\quad
=
\left<
\hat{\mathbf{E}}_{(\text{B})}(\mathbf{r}_a,\omega)
\otimes
\hat{\mathbf{E}}_{(\text{B})}^{\dag}(\mathbf{r}_b,\omega)
\right>
\nonumber \\
&\quad\quad\quad\quad\quad\quad
+
\left<
\hat{\mathbf{E}}_{(\text{M})}(\mathbf{r}_a,\omega)
\otimes
\hat{\mathbf{E}}_{(\text{M})}^{\dag}(\mathbf{r}_b,\omega)
\right>.
\label{eqn:field_correlation}
\end{flalign}
Note that in having \eqref{eqn:field_correlation} there are two cross terms which are
$
\left<
\hat{\mathbf{E}}_{(\text{B})}(\mathbf{r}_a,\omega)
\otimes
\hat{\mathbf{E}}_{(\text{M})}^{\dag}(\mathbf{r}_b,\omega)
\right>
$
and
$
\left<
\hat{\mathbf{E}}_{(\text{M})}(\mathbf{r}_a,\omega)
\otimes
\hat{\mathbf{E}}_{(\text{B})}^{\dag}(\mathbf{r}_b,\omega)
\right>
$.
But these cross terms become zero due to the orthonormal properties of multimode Fock states described in \eqref{eqn:Ortho_Fock_3}.

We shall check the validity of the modified Langevin noise formalism by backsubstituting the electric BA/MA field operators in \eqref{eqn:BA/MA_fields} into \eqref{eqn:SF_field_correlation} to see if it retrieves the original definition of spectral functions in \eqref{eqn:spectral_function}.
If so, we can say that the modified Langevin noise formalism with BA/MA fields makes an open/lossy EM system in thermal equilibrium at every single point in $V$.

Backsubstituting \eqref{eqn:BA/MA_fields} into \eqref{eqn:SF_field_correlation}, we can rewrite the field correlation function in terms of BA/MA fields by
\begin{flalign}
&\left<
\hat{\mathbf{E}}(\mathbf{r}_a,\omega)
\otimes
\hat{\mathbf{E}}^{\dag}(\mathbf{r}_b,\omega)
\right>
=
\left<
\hat{\mathbf{E}}_{(\text{B})}(\mathbf{r}_a,\omega)
\otimes
\hat{\mathbf{E}}_{(\text{B})}^{\dag}(\mathbf{r}_b,\omega)
\right>
\nonumber \\
&\quad\quad\quad\quad\quad\quad
+
\left<
\hat{\mathbf{E}}_{(\text{M})}(\mathbf{r}_a,\omega)
\otimes
\hat{\mathbf{E}}_{(\text{M})}^{\dag}(\mathbf{r}_b,\omega)
\right>
\label{eqn:field_correlation}
\end{flalign}
where $\left<\hat{O}\right>\triangleq\mel{\left\{0\right\}}{\hat{O}}{\left\{0\right\}}$ for an operator $\hat{O}$, and it applies if not specified.
Note that in having \eqref{eqn:field_correlation} there are two cross terms which are
$
\left<
\hat{\mathbf{E}}_{(\text{B})}(\mathbf{r}_a,\omega)
\otimes
\hat{\mathbf{E}}_{(\text{M})}^{\dag}(\mathbf{r}_b,\omega)
\right>
$
and
$
\left<
\hat{\mathbf{E}}_{(\text{M})}(\mathbf{r}_a,\omega)
\otimes
\hat{\mathbf{E}}_{(\text{B})}^{\dag}(\mathbf{r}_b,\omega)
\right>
$.
But these become zero due to the orthonormal properties of multimode Fock states described in \eqref{eqn:Ortho_Fock_3}.
Substituting \eqref{eqn:MA_field} into the second term on the RHS of \eqref{eqn:field_correlation} yields
\begin{flalign}
&
\left<
\hat{\mathbf{E}}_{(\text{M})}(\mathbf{r}_a,\omega)
\otimes
\hat{\mathbf{E}}_{(\text{M})}^{\dag}(\mathbf{r}_b,\omega)
\right>
=
\frac{\hbar\omega^4}{\pi \epsilon_0 c^{4}}
\times
\nonumber \\
&\quad\quad\quad
\iiint_{V_{m}}d\mathbf{r}
\chi_{I}(\mathbf{r},\omega)
\overline{\mathbf{G}}(\mathbf{r}_a,\mathbf{r},\omega)
\cdot
\overline{\mathbf{G}}^{*}(\mathbf{r},\mathbf{r}_b,\omega).
\label{eqn:FC_MA}
\end{flalign}
Based on the dyadic-dyadic Green theorem \cite{Drezet2017Equivalence}, the integral in \eqref{eqn:FC_MA} can be evaluated by
\begin{flalign}
&
\iiint_{V_{m}}d\mathbf{r}'
\epsilon_I(\mathbf{r}',\omega)
\overline{\mathbf{G}}(\mathbf{r}_{a},\mathbf{r}',\omega)
\cdot
\overline{\mathbf{G}}^{*}(\mathbf{r}',\mathbf{r}_{b},\omega)
=
\frac{c^2}{\omega^2}
\times
\nonumber \\
&\quad\quad\quad\quad\quad\quad
\Bigl(
\Im\left\{
\overline{\mathbf{G}}(\mathbf{r}_{a},\mathbf{r}_{b},\omega)
\right\}
-
\overline{\boldsymbol{\mathcal{F}}}(\mathbf{r}_{a},\mathbf{r}_{b},\omega)
\Bigr)
\label{eqn:DDGT}
\end{flalign}
where 
\begin{flalign}
&
\overline{\boldsymbol{\mathcal{F}}}(\mathbf{r}_a,\mathbf{r}_b,\omega)
=
\frac{\omega \sqrt{\epsilon_0}}{c}\times
\nonumber \\
&\quad\quad\quad
\oiint_{S_{\infty}}d\mathbf{r}
\overline{\mathbf{G}}(\mathbf{r}_a,\mathbf{r},\omega)
\cdot
\hat{n}\times\hat{n}\times
\overline{\mathbf{G}}^{*}(\mathbf{r},\mathbf{r}_b,\omega),
\end{flalign}
$\hat{n}$ denotes the outward normal vector on $S_{\infty}$.
Therefore, one can rewrite \eqref{eqn:FC_MA} using the dyadic-dyadic Green theorem by
\begin{flalign}
&
\left<
\hat{\mathbf{E}}_{(\text{M})}(\mathbf{r}_a,\omega)
\otimes
\hat{\mathbf{E}}_{(\text{M})}^{\dag}(\mathbf{r}_b,\omega)
\right>
=
\frac{\hbar\omega^2 \mu_0}{\pi}
\times
\nonumber \\
&\quad\quad\quad\quad\quad\quad
\Bigl(
\Im\left\{
\overline{\mathbf{G}}(\mathbf{r}_{a},\mathbf{r}_{b},\omega)
\right\}
-
\overline{\boldsymbol{\mathcal{F}}}(\mathbf{r}_{a},\mathbf{r}_{b},\omega)
\Bigr).
\label{eqn:FC_MA}
\end{flalign}
We now evaluate the first term on the RHS of \eqref{eqn:field_correlation} with the substitution of \eqref{eqn:BA_field} by
\begin{flalign}
&
\left<
\hat{\mathbf{E}}_{(\text{B})}(\mathbf{r}_a,\omega)
\otimes
\hat{\mathbf{E}}_{(\text{B})}^{\dag}(\mathbf{r}_b,\omega)
\right>
=
\frac{\hbar\omega}{2 \left(2 \pi\right)^3}
\oiint_{S_{k}} d\mathbf{k}
\nonumber \\
&\quad\quad
\sum_{{\lambda}\in\left\{1,2\right\}}
\boldsymbol{\Phi}_{(\text{tot})}(\mathbf{r}_a,\mathbf{k},\lambda,\omega)
\otimes
\boldsymbol{\Phi}^{*}_{(\text{tot})}(\mathbf{r}_b,\mathbf{k},\lambda,\omega).
\end{flalign}
Eventually, one can obtain the following expression for the field correlation function
\begin{flalign}
&\left<
\hat{\mathbf{E}}(\mathbf{r}_a,\omega)
\otimes
\hat{\mathbf{E}}^{\dag}(\mathbf{r}_b,\omega)
\right>
= 
\frac{\hbar\omega^2\mu_0}{\pi}
\Biggl(
\Im\left\{
\overline{\mathbf{G}}(\mathbf{r}_a,\mathbf{r}_b,\omega)
\right\}
\nonumber\\
&
\quad
-\overline{\boldsymbol{\mathcal{F}}}(\mathbf{r}_a,\mathbf{r}_b,\omega)
+
\frac{\pi}{2\omega\mu_0}
\frac{1}{\left(2\pi\right)^{3}}
\oiint_{S_{k}} d\mathbf{k}
\nonumber\\
&
\quad
\sum_{{\lambda}\in\left\{1,2\right\}}
\boldsymbol{\Phi}_{(\text{tot})}(\mathbf{r}_a,\mathbf{k},\lambda,\omega)
\otimes
\boldsymbol{\Phi}^{*}_{(\text{tot})}(\mathbf{r}_b,\mathbf{k},\lambda,\omega)
\Biggr).
\end{flalign}
Hence, if the following condition
\begin{flalign}
&
\overline{\boldsymbol{\mathcal{F}}}(\mathbf{r}_a,\mathbf{r}_b,\omega)
=\frac{\pi}{2\omega\mu_0\left({2\pi}\right)^{3}}
\oiint_{S_{k}} d\mathbf{k}
\nonumber \\
&\quad
\sum_{{\lambda}\in\left\{1,2\right\}}
\boldsymbol{\Phi}_{(\text{tot})}(\mathbf{r}_a,\mathbf{k},\lambda,\omega)
\otimes
\boldsymbol{\Phi}^{*}_{(\text{tot})}(\mathbf{r}_b,\mathbf{k},\lambda,\omega)
\label{eqn:thermal_eq_cond}
\end{flalign}
would hold, one can retrieve the original definition of the spectral function from the field correlation function substituted by the BA/MA fields, i.e.,
\begin{flalign}
&\left<
\hat{\mathbf{E}}(\mathbf{r}_a,\omega)
\otimes
\hat{\mathbf{E}}^{\dag}(\mathbf{r}_b,\omega)
\right>
=
\frac{\hbar\omega^2\mu_0}{\pi}
\Im\left\{
\overline{\mathbf{G}}(\mathbf{r}_a,\mathbf{r}_b,\omega)
\right\}
\nonumber \\
&\quad\quad\quad\quad\quad\quad\quad\quad\quad\quad
=
-
\frac{\hbar\omega^2 \mu_0 }{2\pi}
\overline{\mathbf{A}}_{0}(\mathbf{r}_a,\mathbf{r}_b,\omega).
\end{flalign}
In other words, the consideration of both BA/MA fields can only ensure the system to be in thermal equilibrium.
We will numerically validate the thermal equilibrium condition \eqref{eqn:thermal_eq_cond} by considering a spontaneous emission rate of a two-level atom located at the inside or outside of a lossy dielectric slab.

\section{Simulation results: \\Purcell factor of a two-level atom located either inside or outside \\a lossy dielectric slab}\label{sec_VI}
\subsection{Problem description}
\begin{figure}
\centering
\includegraphics[width=\linewidth]{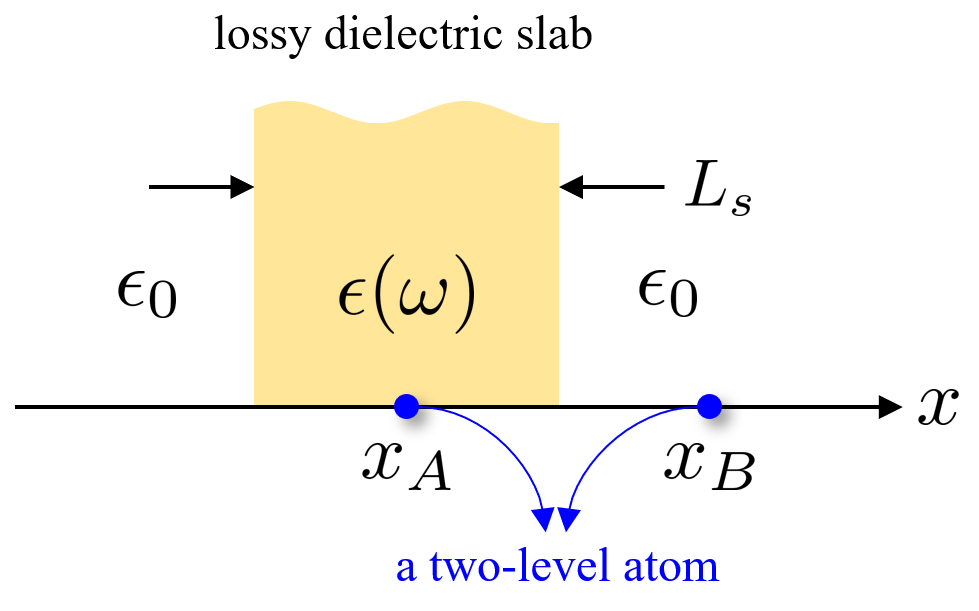}
\caption{Analyzing the spontaneous emission rate when a two-level atom (TLA) is located either inside or outside a lossy dielectric slab.}
\label{fig:TLA_ADI_slab_problem_geometry}
\end{figure}

In this section, we present simulation results of the spontaneous emission rate (SER) of a two-level atom (TLA) when the TLA is located either inside or outside a lossy dielectric slab.
Fig. \ref{fig:TLA_ADI_slab_problem_geometry} illustrates the relevant problem geometry.

\begin{figure}
\centering
\includegraphics[width=\linewidth]{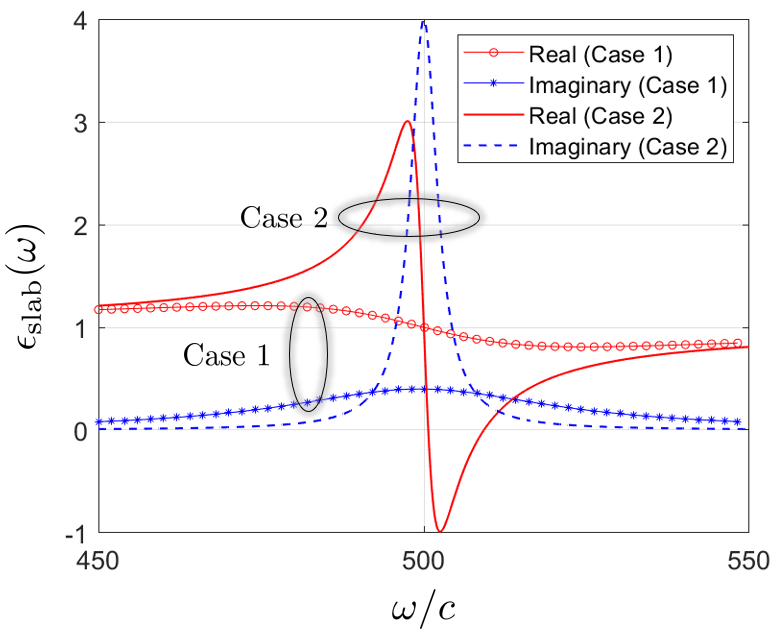}
\caption{Real and imaginary parts of the electric susceptibility $\chi(\omega)$ of the lossy dielectric slab for Case 1 ($\gamma=50$) and Case 2 ($\gamma=5$).}
\label{fig:medium_parameter}
\end{figure}

The lossy dielectric slab is assumed to be spatially homogeneous and have the following electric susceptibility
\begin{flalign}
\chi(x,\omega)
=
\chi(\omega)
=
\left\{
\begin{matrix}
\frac{\omega_p^2}{\omega_0^{2}-\omega^{2}+i\omega \gamma} & \text{for } -\frac{L_{s}}{2} \leq x\leq \frac{L_{s}}{2} \\
0 & \text{elsewhere}
\end{matrix}
\right.
\nonumber
\end{flalign}
where thickness of the slab $L_{s}=62.5$ [mm], $\omega_p=100c$, and $\omega_0=500c$.
We consider two different loss factors $\gamma=50$ (high loss for Case 1) and $\gamma=5$ (low loss for Case 2).
Fig. \ref{fig:medium_parameter} compares real and imaginary parts of the electric susceptibility of Case 1 and 2.
It can be observed in Fig. \ref{fig:medium_parameter} that $\text{Im}\left(\chi_{\text{slab}}\right)$ (equivalently, dielectric medium loss) for both cases becomes maximized at $\omega=500c$ but different quality factors.

Assume that a TLA is located at $x_a$ with a transition frequency $\omega_{a}$.
We consider two different TLA's locations, i.e., $x_a=x_A=0$ (inside the slab for Case A) and $x_a=x_B=L_s$ (outside the slab for Case B), as illustrated in Fig. \ref{fig:TLA_ADI_slab_problem_geometry}.

\subsection{Spontaneous emission rate based on the Fermi-Golden rule}
In this one-dimensional simulation setup, we shall assume that electric field operators are polarized along $y$-axis.
One can then evaluate the SER, denoted by $\Gamma$, based on the Fermi-Golden rule \cite{fox2006quantum} below:
\begin{flalign}
&
\Gamma(\omega_a)
=
\frac{2\pi}{\hbar^2}
\left|
\mathbf{d}
\cdot
\hat{y}
\right|^{2}
\int_{0}^{\infty}d\omega
\delta(\omega_{a}-\omega)
\times
\nonumber \\
&\quad\quad\quad\quad\quad\quad\quad\quad
\mel
{\left\{0\right\}}
{
\hat{E}_{y}(x_{a},\omega)
\hat{E}_{y}^{\dag}(x_{a},\omega)
}
{\left\{0\right\}}
\label{eqn:SER_TLA_ADI_slab}
\end{flalign}
where $\mathbf{d}$ is a dipole moment of a TLA.

Based on the Fermi-Golden rule, we evaluate the SERs using four different methods.

\subsubsection{Method 1: Spectral Function Approach (SFA)}
The field correlation in the SER expression \eqref{eqn:SER_TLA_ADI_slab} can be related by the spectral function.
And the spectral function takes the imaginary part of the Green's function (see \eqref{eqn:spectral_function}).
Thus, based on the SFA the SER takes the following form:
\begin{flalign}
\Gamma^{(\text{M1})}(\omega_a)
=
\frac{2\omega_a^2 \left|\mathbf{d}\cdot\hat{y}\right|^{2}}{\hbar c^2 \epsilon_0 }
\Im \left\{G(x_{a},x_{a},\omega_a)\right\}
\label{eqn:FDT_SER}
\end{flalign}
where the Green's function is governed by
\begin{flalign}
\Bigl(\frac{d^2}{dx^2}+\frac{\omega^2}{c^2}\epsilon_r(x,\omega)\Bigr) G(x,x',\omega)
=
-
\delta(x-x').
\label{eqn:1D_GF}
\end{flalign}

Note that the formulation \eqref{eqn:FDT_SER} is also widely used to estimate the SER of a TLA in quantum optics.
In this work, we have performed one-dimensional FEM numerical simulations of point-source-radiation problems in \eqref{eqn:1D_GF} to evaluate the SER in \eqref{eqn:FDT_SER}.

\subsubsection{Method 2: Second quantization of the microscopic model via numerical mode decomposition}
As another ground truth, we calculate the SER based on the second quantization of the microscopic model via numerical mode decomposition.
In the microscopic model, instead of employing the effective permittivity $\epsilon(\omega)$ for the lossy slab, reservoir oscillator fields are introduced over the slab region and interact with EM fields.
Their explicit interactions can model the EM energy loss by the lossy slab.
Furthermore, random initial conditions of the reservoir oscillator fields are associated with Langevin noise current sources. 

Based on the microscopic model \cite{Philbin2010canonical,Sha2018Dissipative}, one can formulate the continuum generalized Hermitian eigenvalue problem (GH-EVP) \cite{Dorier2019Canonical,Na2021Diagonalization}.
We can extract eigenmodes from the GH-EVP to perform the second quantization of the microscopic model.
However, it is difficult to account for the infinite degrees of freedom of the reservoid oscillator fields when extracting numerical eigenmodes. 
In this work, we properly coarse-grained reservoir oscillator fields over both the slab region and resonant frequency domain.

Using extracted numerical eigenmodes, one can represent an electric field operator by
\begin{flalign}
\hat{E}_{y}^{(+)}(x_a,t)
\approx
i\sum_{m} \tilde{E}_{m}(x_{a})\sqrt{\frac{\hbar\omega_{m}}{2}}e^{-i\omega_m t}\hat{c}_{m}
\label{eqn:E_field_exp_model}
\end{flalign}
where, for $m$-th numerical eigenmode in the above series, $\tilde{E}_{m}(x)$ denotes the electric field part of the eigenmode, $\omega_{m}$ is eigenfrequency, $\hat{c}_{m}$ ($\hat{c}_{m}^{\dag}$) is an annihilation (creation) operator satisfying the standard bosonic commutator relations.
Note that, with numerical eigenmodes, one can formally represent the Hamiltonian operator by
\begin{flalign}
\hat{H}=\sum_{m}\hbar\omega_{m} \hat{c}^{\dag}_{m}\hat{c}_{m}.
\end{flalign}

By substituting \eqref{eqn:E_field_exp_model} into \eqref{eqn:SER_TLA_ADI_slab}, the SER can be calculated by
\begin{flalign}
\Gamma^{(\text{M2})}(\omega_a)
\approx
\frac{\left|
\mathbf{d}
\cdot
\hat{y}
\right|^{2}}{\hbar}
\sum_{m}
\frac{\eta \omega_m
\tilde{E}_{m}^{*}(x_{a})
\tilde{E}_{m}(x_{a})}{(\omega_a-\omega_{m})^2+\eta^2}
\end{flalign}
for a small $\eta$ where the delta function is approximated by \cite{Chew2019Green}
\begin{flalign}
\delta(\omega_a-\omega)
=
\frac{1}{\pi}
\lim_{\eta \to 0}
\frac{\eta}{(\omega_a-\omega)^2+\eta^2}.
\end{flalign}
The reason why the delta function was approximated by the above bandpass filter is due to the fact that numerical eigenmodes form a countably-finite eigenspectrum owing to the discretization.
The quality factor of the bandpass filter can be controlled by $\eta$, i.e., the smaller $\eta$ is, the higher the quality factor is.
With a larger problem domain size, finer mesh, increasing the extent of coarse-graining bath oscillators, one can have denser eigenfrequencies so that $\eta$ can be much smaller converging to a real delta function.

\subsubsection{Method 3: Modified Langevin noise formalism}
Now, we evaluate the SER based on the modified Langevin noise formalism, i.e., substituting expressions for BA/MA fields in \eqref{eqn:BA/MA_fields} into \eqref{eqn:SER_TLA_ADI_slab}.
For this one-dimensional case, the monochromatic electric field operator \eqref{eqn:BA/MA_fields} can be simplified into
\begin{flalign}
&\hat{E}_{y}(x_a,\omega)
=
\frac{i}{2}\sqrt{\frac{\hbar\omega}{\pi}}
\sum_{k_{x}\in\left\{\pm k\right\}}
\Phi_{(\text{tot})}(x_a,k_{x},\omega)
\hat{a}(k_{x},\omega)
\nonumber \\
&
\!
+
\frac{i\omega^2}{c^2}
\int_{-\frac{L_s}{2}}^{\frac{L_s}{2}} dx'
\sqrt{\frac{\hbar\chi_{I}(x',\omega)}{\pi\epsilon_0}}
G(x_a,x',\omega)
\hat{f}(x',\hat{y},\omega)
\label{eqn:BA/MA_e_field_1D}
\end{flalign}
where wavenumber in free space $k=\omega/c$.
Substituting the above into \eqref{eqn:SER_TLA_ADI_slab} and using the bosonic commutators in \eqref{eqn:BC_a} and \eqref{eqn:BC_f}, one can have the following expression for the SER:
\begin{flalign}
\Gamma^{(\text{modified LN})}(\omega_a)=\Gamma^{(\text{B})}(\omega_a)+\Gamma^{(\text{M})}(\omega_a)
\label{eqn:SER_BA/MA}
\end{flalign}
where $\Gamma^{(\text{M})}$ and $\Gamma^{(\text{M})}$ are the SER by BA and MA fields, respectively, taking the form of
\begin{flalign}
&\Gamma^{(\text{B})}(\omega_a)
=
\frac{\omega_a\left|
\mathbf{d}
\cdot
\hat{y}
\right|^{2}}{2\hbar}
\sum_{k_{x}\in\left\{\pm k\right\}}
\left|
\Phi_{(\text{tot})}(x_a,k_{x},\omega)
\right|^{2}
\\
&\Gamma^{(\text{M})}(\omega_a)
=
\frac{2 \omega_{a}^2\left|
\mathbf{d}
\cdot
\hat{y}
\right|^{2}}
{\hbar c^2 \epsilon_0}
\Biggl(
\frac{\omega_a^{2}}{c^{2}}
\int_{-\frac{L_s}{2}}^{\frac{L_s}{2}} dx'
\chi_{I}(x',\omega_a)
\nonumber \\
&\quad\quad\quad\quad\quad\quad\quad\quad
\times
G(x_a,x',\omega_a)
G^{*}(x',x_a,\omega_a)
\Biggr).
\label{eqn:SER_BA/MA}
\end{flalign}
Note that we performed one-dimensional FEM numerical simulations of plane-wave-scattering and point-source-radiation problems to obtain $\Phi_{(\text{tot})}(x_a,k_{x},\omega)$ and $G(x_a,x',\omega_a)$, respectively.

\subsubsection{Method 4: Original Langevin noise formalism}
The original Langevin noise formalism accounts for MA fields only, which are the second term on the RHS of \eqref{eqn:BA/MA_e_field_1D}.
Thus, it can be easily shown that when using the original Langevin noise formalism the SER is the same as $\Gamma^{(\text{M})}(\omega_a)$ in \eqref{eqn:SER_BA/MA}:
\begin{equation} 
\begin{aligned}
&\Gamma^{(\text{original LN})}(\omega_a)=
\Gamma^{(\text{M})}(\omega_a)
\end{aligned}
\label{eqn:Welsch_SER}
\end{equation}

\subsection{Comparison of Purcell factors calculated by four different methods}
Since the one-dimensional free space SER is given by
\begin{flalign}
\Gamma_0(\omega_a)
=
\frac{\omega_{a}\left|\mathbf{d}\cdot\hat{y}\right|^{2}}{\hbar \epsilon_0 c},
\label{eqn:Case_2}
\end{flalign}
the Purcell factor can be evaluated by
\begin{flalign}
\text{Purcell factor}(\omega_a)= \Gamma(\omega_a)/\Gamma_0(\omega_a).
\end{flalign}

We now compare Purcell factors calculated by using Methods 1, 2, 3, and 4 for two different loss factors (Case 1 and Case 2) and two different locations of TLA (Case A and Case B).
In total, four possible cases are labeled by 1-A, 1-B, 2-A, and 2-B.
\begin{figure*}
\centering
\subfloat[Case 1-A ($\gamma=50$ and $x_a=x_A=0$)]
{\includegraphics[width=.5\linewidth]{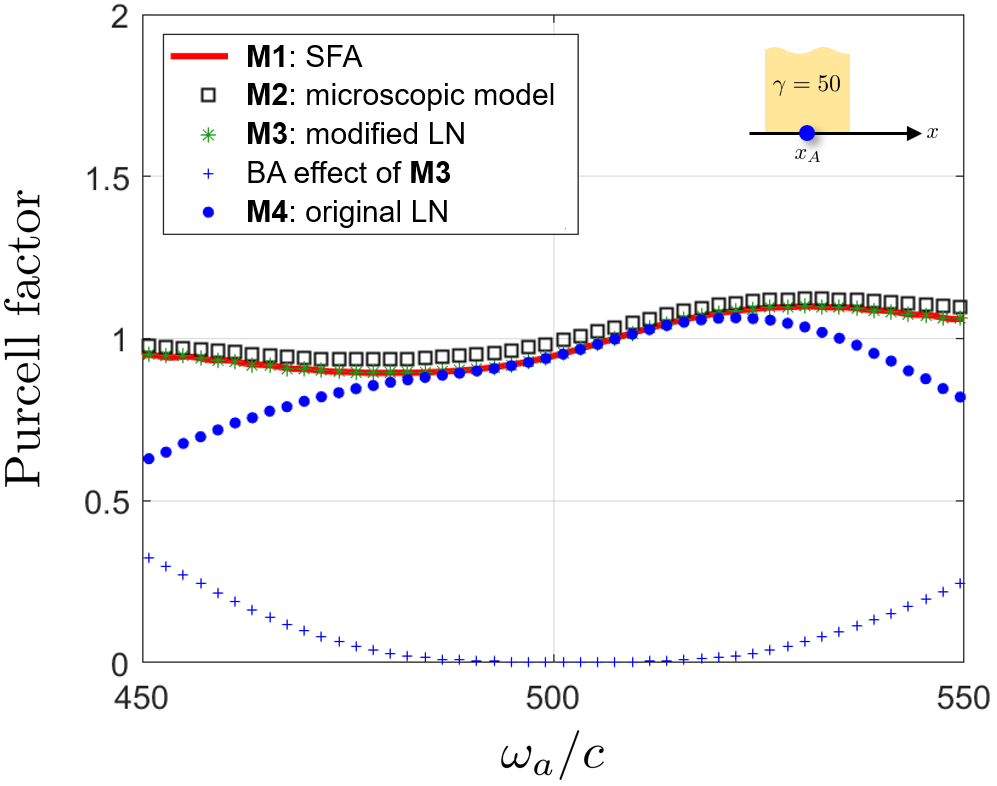}
\label{fig:case_D}
}
\subfloat[Case 1-B ($\gamma=50$ and $x_a=x_B=L_s$)]
{\includegraphics[width=.5\linewidth]{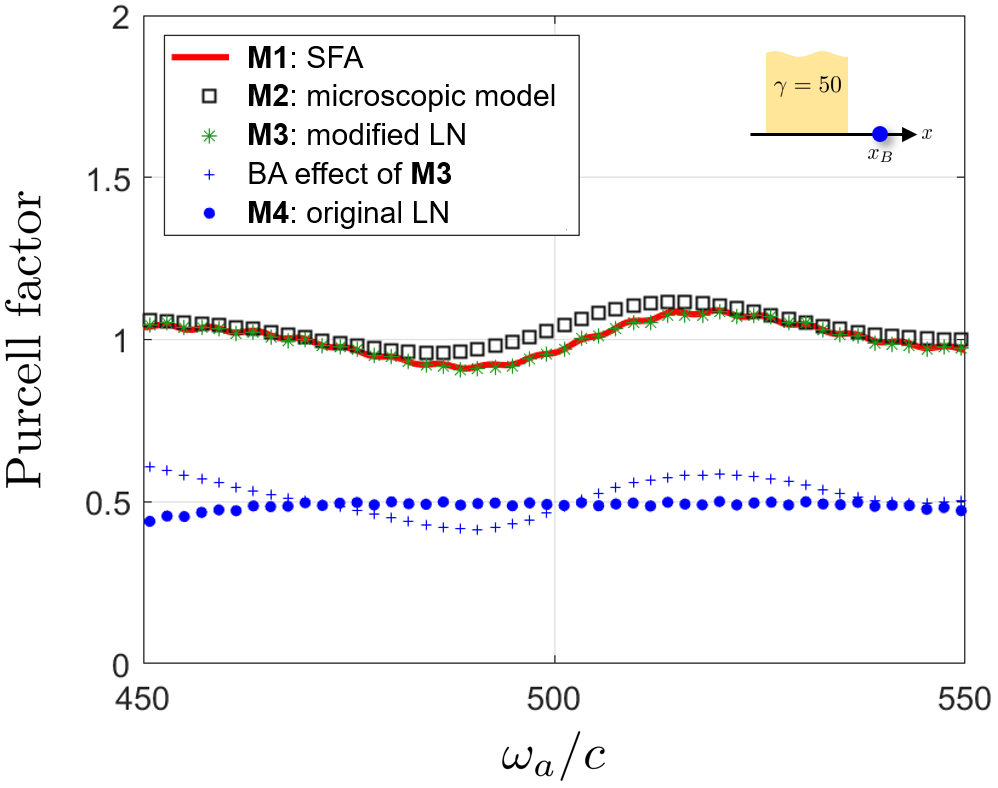}
\label{fig:case_C}
}
\caption{Purcell factors of a two-level atom versus an atomic transition frequency $\omega_a$ for Case 1 (lossy factor $\Gamma=50$, i.e., high loss) and (a) $x_a=x_A=0$ (inside the lossy slab) and (b) $x_a=x_B=L_s$ (outside the lossy slab). Note that LN is the abbreviation of Langevin noise.}
\label{fig:SER_comp_Case_1}
\end{figure*}
\begin{figure*}
\centering
\subfloat[Case (2-A): $x_a=x_A=0$ and $\gamma=5$]
{\includegraphics[width=.5\linewidth]{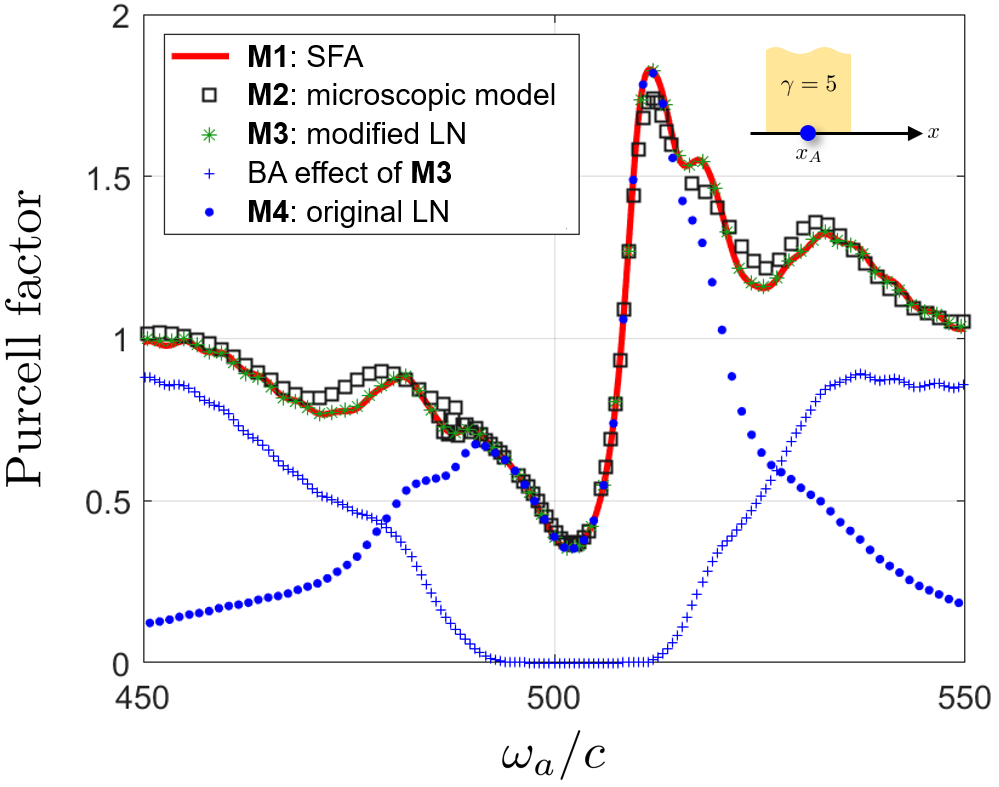}
\label{fig:case_B}
}
\subfloat[Case (2-B): $x_a=x_B=L_s$ and $\gamma=5$]
{\includegraphics[width=.5\linewidth]{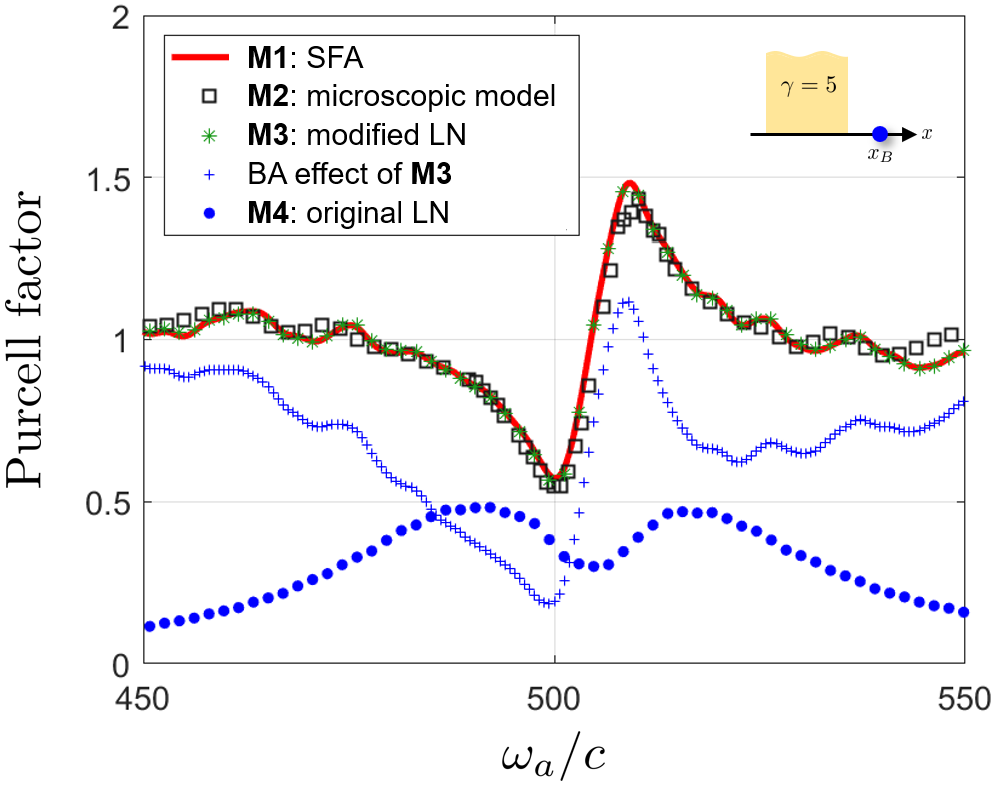}
\label{fig:case_A}
}
\caption{Purcell factors of a two-level atom versus an atomic transition frequency $\omega_a$ for Case 2 (lossy factor $\gamma=5$, i.e., low loss) and (a) $x_a=x_A=0$ (inside the lossy slab) and (b) $x_a=x_B=L_s$ (outside the lossy slab).}
\label{fig:SER_comp_Case_2}
\end{figure*}
Case 1-A and Case 1-B are illustrated in Fig. \ref{fig:case_D} and Fig. \ref{fig:case_C}, respectively.
First of all, it is observed that the modified Langevin noise formalism (Method 3 illustrated by green $*$ markers) has an excellent agreement with the two reference cases whereas the original Langevin noise formalism exhibits significant deviations from them in general.
This numerical experiment proves the validity of the modified Langevin noise formalism.

Let us further observe BA and MA contributions separately.
When the TLA is located inside the slab (Case 1-A), MA contributions become dominant as the loss of the slab is maximized around $\omega_a \approx 500c$ (see Fig. \ref{fig:medium_parameter}).
On the other hand, BA contributions are subtle around $\omega_a\approx 500c$.
This can be explained as follows:
Incident plane waves, which produce BA fields, cannot penetrate deep into the lossy slab and reach the TLS's location.
Consequently, BA fields would have extremely small contributions to the formation of local density of states (LDOS) at the TLS's location.
On the other hand, Langevin noise current operators \eqref{eqn:Langevin_Noise_Current_Operator_PLN} are proportional to the loss of the dielectric medium; therefore, the higher medium loss, the stronger MA fields can be produced.
These strong MA (near) fields would contribute to the formation of LDOS at the TLS's location.\footnote{The higher loss makes the propagation of MA fields quickly attenuated.}
This may explain why the use of the previous LN model considering effects of MA fields only was so popular in quantum optics, especially, when a TLA is buried deep inside lossy dielectric objects or some places where BA fields barely affect the formation of LDOSs.

Consider now Case 1-B (see Fig. \ref{fig:case_C}) where the TLA is now located outside the slab.
BA fields now start having contributions to the net Purcell factors.
We can further deduce that  the formation of LDOSs is mainly contributed by (i) MA fields escaping from the slab toward the right and (ii) BA fields whose incident plane waves coming from the right side.
From these observations, we can figure out that both BA and MA fields should be taken into account on an equal footing in general cases, especially, BA fields can affect the formation of LDOSs at the TLS's location.

Simulation results for Case 2 (loss factor $\gamma=5$) are illustrated in Fig. \ref{fig:SER_comp_Case_2} for two two TLA's locations again.
Similar tendency can be observed that (i) when the TLA is buried inside the lossy slab, MA effects are dominant, and (ii) when the TLA is outside the slab, both BA and MA fields contribute to the net SER.

\subsection{Numerical validation of thermal equilibrium condition \eqref{eqn:thermal_eq_cond}}
\begin{figure*}
\centering
\subfloat[Case 1 ($\gamma=50$)]
{\includegraphics[width=.5\linewidth]{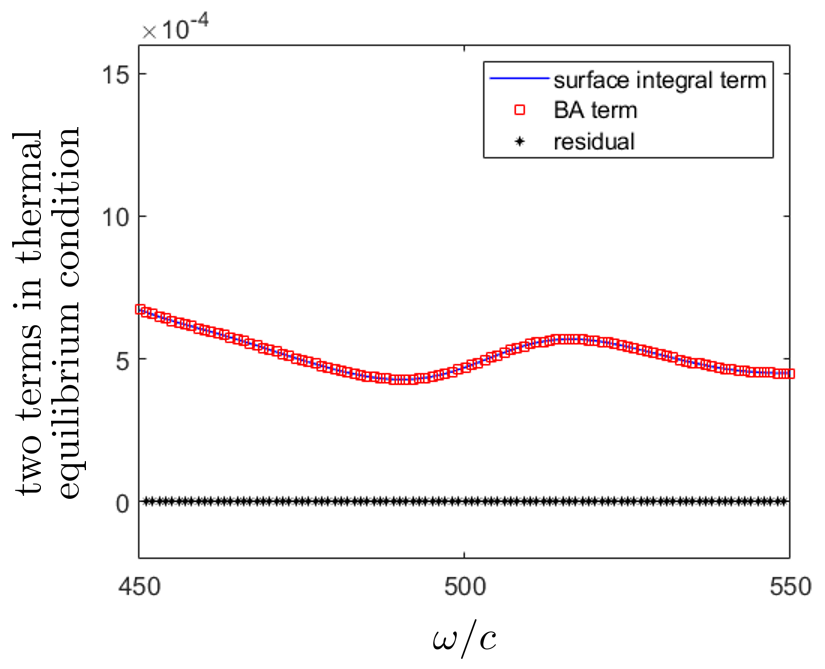}
\label{fig:TEC_HL_Case_C}
}
\subfloat[Case 2  ($\gamma=5$)]
{\includegraphics[width=.5\linewidth]{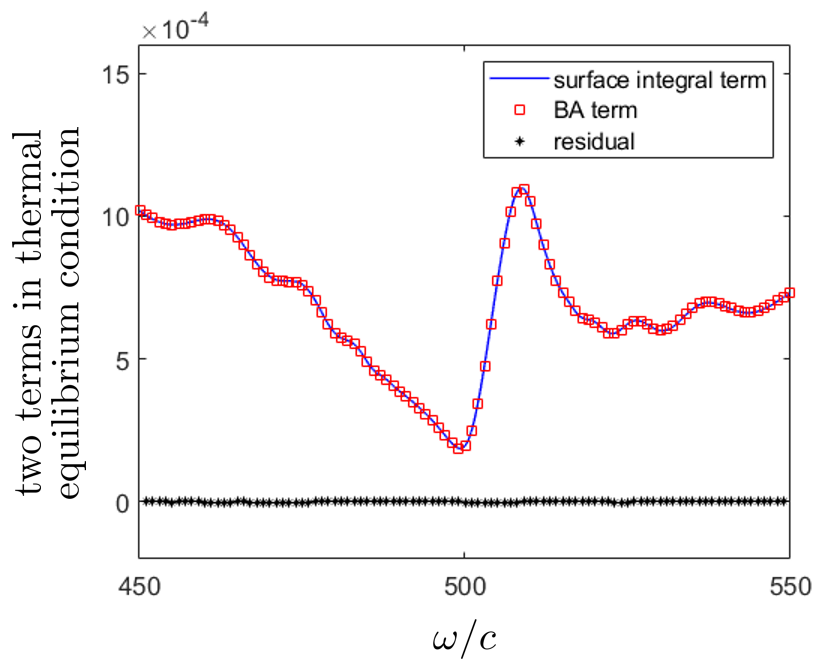}
\label{fig:TEC_LL_Case_C}
}
\caption{Numerical validation of the thermal equilibrium condition in \eqref{eqn:1D_TEC} for Case 1 and 2 where $x_a=x_b=x_B=L_s$. The surface integral term (LHS in \eqref{eqn:1D_TEC}) is evaluated by using the analytic expression given in \cite{Dorier2020Critical} whereas the BA term (RHS in \eqref{eqn:1D_TEC}) is numerically calculated by FEM simulation.}
\label{fig:thermal_eq_cond_comp_num_res}
\end{figure*}

Here, we numerically validate the thermal equilibrium condition \eqref{eqn:thermal_eq_cond}.
For the one-dimensional case, we can simplify the thermal equilibrium condition into
\begin{widetext}
\begin{flalign}
\underbrace{
\mathcal{F}(x_\alpha,x_\beta,\omega)
}_{\text{surface integral term}}
=
\underbrace{
\frac{1}{4\omega\mu_0}
\sum_{k_{x}=\pm\frac{\omega_a}{c}}
\Phi_{(\text{tot})}(x_\alpha,k_{x},\omega)
\Phi^{*}_{(\text{tot})}(x_\beta,k_{x},\omega)
}_{\text{BA term}}
\label{eqn:1D_TEC}
\end{flalign}
\end{widetext}
The LHS (the surface integral term) can be analytically calculated by the formula \cite{Dorier2020Critical}.
We numerically evaluate the RHS (BA term) using the FEM simulations and compare two terms.
We assume that $x_\alpha=x_\beta=x_B=L_s$.
The results for Case 1 and Case 2 are illustrated in Fig. \ref{fig:thermal_eq_cond_comp_num_res}.
It can be observed that the two terms are almost same with subtle residuals.
Hence, our numerical experiment validates the thermal equilibrium condition, which strongly supports that BA and MA fields together can make open and lossy EM systems quasi-Hermitian or in the thermal equilibrium. 

\section{Summary and Concluding Remarks}
We have proposed the numerical framework by incoporating the use of numerical methods into the modified Langevin formalism with boundary-assisted (BA) and medium-assisted (MA) fields for quantization of electromagnetic systems involving both radiation and dielectric losses.
For this demonstraction, we have used the finite element method to solve plane-wave-scattering and point-source-radiation problems for obtaining BA/MA fields, respectively.
But other computational electromagnetic methods are also available. 
Importantly, for the first time, we have numerically validated the modified Langevin formalism with BA/MA fields by calculating the spontaneous emission rate of a two-level atom either inside or outside a lossy dielectric slab.
The numerical evaluation of substituting the BA/MA fields into the Fermi-Golden rule in \eqref{eqn:SER_TLA_ADI_slab} agreed with the typical expression for the spontaneous emission rate in \eqref{eqn:FDT_SER}, which is proportional to the imaginary part of the Green's function derivable through the spectral function approach.
Our observation indicates that the consideration of BA fields is essential whenever the radiation loss is present, for example, finite-sized lossy dielectrics.
The proposed numerical framework for the modified Langevin noise formalism with BA/MA fields can be utilized for modeling  arbitrary quantized lossy electromagnetic systems and quantification of various practical quantum optics problems associated with plasmonic structures, metasurfaces, and nanoparticles.
It should be mentioned that the proposed framework can model the expectation value of arbitrary operators or observables (e.g., higher-order correlation) with respect to various initial quantum states (e.g., entangled states). 
This cannot be done by using spectral function approach which can only provide the first-order correlation for thermal or ground states.

\begin{acknowledgments}
The work is funded by NSF 1818910 award and a startup fund at Purdue university.
\end{acknowledgments}

\appendix
\section{Operator-form dyadic-dyadic Green theorem}
One can formally prove the dyadic-dyadic Green theorem \eqref{eqn:DDGT}, as shown by our recent work \cite{Chew2019Green}, by writing \eqref{eqn:DGF} in an operator form as
\begin{flalign}
\left(\hat{\mathcal{S}}-\frac{\omega^2}{c^2}\hat{\mathcal{M}}\right)\hat{\mathcal{G}}=\hat{\mathcal{I}}
\label{eqn:DGF_op}
\end{flalign}
where $\hat{\mathcal{S}}$ and $\hat{\mathcal{M}}$ are operator forms of $\nabla\times\mu_{r}^{-1}(\mathbf{r})\nabla\times$ and $\epsilon_r(\mathbf{r},\omega)$, respectively, $\hat{\mathcal{G}}$ is an operator form of the dyadic Green's function $\overline{\mathbf{G}}(\mathbf{r},\mathbf{r}',\omega)$, and $\hat{\mathcal{I}}$ is an operator form of $\delta(\mathbf{r}-\mathbf{r}')\overline{\mathbf{I}}$.
When $\left(\hat{\mathcal{S}}-\frac{\omega^2}{c^2}\hat{\mathcal{M}}\right)$ is non-singular or invertible,  
\begin{flalign}
\hat{\mathcal{G}}^{-1}=\hat{\mathcal{S}}-\frac{\omega^2}{c^2}\hat{\mathcal{M}}.
\end{flalign}
Thus, one can derive an interesting expression for a spectral function operator $\hat{\mathcal{A}}$ similar to \cite{Chew2019Green,datta_2005} such as
\begin{flalign}
\hat{\mathcal{A}}^{-1}
&= i\left[ \left(\hat{\mathcal{G}}^{a}\right)^{-1} - \hat{\mathcal{G}}^{-1} \right]
\nonumber \\
&= i \left[ 
\left(\hat{\mathcal{S}}^{a}-\hat{\mathcal{S}}\right) 
- \frac{\omega^2}{c^2} 
\left(\hat{\mathcal{M}}^{a}-\hat{\mathcal{M}}\right) \right]
\nonumber \\
&=2 \text{Im}\left(\hat{\mathcal{S}}\right) - 2 \frac{\omega^2}{c^2}\text{Im}\left(\hat{\mathcal{M}}\right)
\label{eqn:Gamma}
\end{flalign}
where superscript $a$ denotes the adjoint operator.
It should be emphasized that 
\begin{flalign}
\text{Im}\left(\hat{\mathcal{S}}\right)\neq 0
\end{flalign}
due to open boundary conditions even if there is no magnetic loss.
Multiplying \eqref{eqn:Gamma} by $\hat{\mathcal{G}}$ and $\hat{\mathcal{G}}^{a}$ from the left and right, respectively, one can have
\begin{flalign}
i\left[ \hat{\mathcal{G}} - \hat{\mathcal{G}}^{a} \right]
=
2 \hat{\mathcal{G}}~\text{Im}\left(\hat{\mathcal{S}}\right)\hat{\mathcal{G}}^{a} 
- 2 \frac{\omega^2}{c^2} \hat{\mathcal{G}}~\text{Im}\left(\hat{\mathcal{M}}\right)\hat{\mathcal{G}}^{a}.
\end{flalign}
Since $i\left[ \hat{\mathcal{G}} - \hat{\mathcal{G}}^{a} \right] =-2 \text{Im}\left(\hat{\mathcal{G}}\right)$, one can retrieve \eqref{eqn:DDGT} in an operator form as
\begin{flalign}
\text{Im}\left(\hat{\mathcal{G}}\right)
=
-\hat{\mathcal{G}}~\text{Im}\left(\hat{\mathcal{S}}\right)\hat{\mathcal{G}}^{a}
+
\frac{\omega^2}{c^2} \hat{\mathcal{G}}~\text{Im}\left(\hat{\mathcal{M}}\right)\hat{\mathcal{G}}^{a},
\label{eqn:dd_gT}
\end{flalign}
which is the correct dyadic-dyadic Green theorem when dielectric medium and radiation losses are present.

Previous works \cite{Gruner1996Green,Dung1998three} assumed $\text{Im}\left(\hat{S}\right)=0$ such that the identity below
\begin{flalign}
\text{Im}\left(\hat{\mathcal{G}}\right)
=
\frac{\omega^2}{c^2} \hat{\mathcal{G}}~\text{Im}\left(\hat{\mathcal{M}}\right)\hat{\mathcal{G}}^{a}.
\label{eqn:dd_gT_previous}
\end{flalign}
This identity has been then accepted without reasonable arguments nor specification of boundary conditions.
As clearly shown, when radiation boundary conditions (causing radiation losses) are used, \eqref{eqn:dd_gT_previous} should be modified into \eqref{eqn:dd_gT}. 
This also implies that the previous electric field operators in \eqref{eqn:prev_LN} should be modified by including the boundary-assistend fields as in \eqref{eqn:BA/MA_fields}.

\bibliography{sample}

\end{document}